\documentclass[a4paper,fleqn]{cas-dc}

\usepackage[numbers]{natbib}
\usepackage{lineno}
\usepackage{hyperref}
\usepackage{graphicx}
\usepackage{array}
\usepackage{tabularx}
\usepackage{multicol}

\begin{document}

\shorttitle{Deep Residual CNN for Multi-Class Chest Infection Diagnosis}
\shortauthors{Kwon et~al.}


\title [mode = title]{Deep Residual CNN for Multi-Class Chest Infection Diagnosis}

\author[1]{Ryan Donghan Kwon}
\ead{ryankwon@ieee.org}

\author[2]{Dohyun Lim}
\ead{22-094@ksa.hs.kr}

\author[1]{Yoonha Lee}
\ead{yoonha724@gmail.com}

\author[3]{Seung Won Lee}
\ead{lsw9290@gmail.com}

\address[1]{Department of Science and Informatics, Hana Academy Seoul, Eunpyeong-gu, Seoul, 03305 Republic of Korea}
\address[2]{Department of Mathematics and Computer Science, Korea Science Academy of KAIST, Busanjin-gu, Busan, 47162 Republic of Korea}
\address[3]{Department of Precision Medicine, School of Medicine, Sungkyunkwan University, Jangan-gu, Suwon-si, Gyeonggi-do, 16419 Republic of Korea}

\begin{abstract}
The advent of deep learning has significantly propelled the capabilities of automated medical image diagnosis, providing valuable tools and resources in the realm of healthcare and medical diagnostics. This research delves into the development and evaluation of a Deep Residual Convolutional Neural Network (CNN) for the multi-class diagnosis of chest infections, utilizing chest X-ray images. The implemented model, trained and validated on a dataset amalgamated from diverse sources, demonstrated a robust overall accuracy of 93\%. However, nuanced disparities in performance across different classes, particularly Fibrosis, underscored the complexity and challenges inherent in automated medical image diagnosis. The insights derived pave the way for future research, focusing on enhancing the model’s proficiency in classifying conditions that present more subtle and nuanced visual features in the images, as well as optimizing and refining the model architecture and training process. This paper provides a comprehensive exploration into the development, implementation, and evaluation of the model, offering insights and directions for future research and development in the field.
\end{abstract}

\begin{keywords}
Deep Learning \sep Convolutional Neural Networks \sep Medical Image Analysis \sep Chest Infections \sep X-ray Image Classification \sep Residual Learning \sep Multi-class Diagnosis \sep Healthcare Technology \sep Automated Diagnosis \sep Image Processing
\end{keywords}

\maketitle

\section{Introduction}
The advent of the COVID-19 pandemic has brought forth unprecedented challenges to the global healthcare system, particularly in the realm of diagnostic procedures for respiratory infections. The criticality of accurate, timely, and scalable diagnostic solutions has been underscored by the rapid transmission and potentially severe manifestations of the virus. In this context, the utilization of deep learning technologies, especially Convolutional Neural Networks (CNNs), in the analysis and interpretation of medical images has emerged as a pivotal tool in enhancing diagnostic capabilities and efficiencies.

Pneumonia, a life-threatening disease that occurs in the lungs due to bacterial or viral infections, has been a focal point in the application of deep learning in medical diagnostics. The disease can be life-endangering if not diagnosed and treated promptly, making early detection vital. A study by Rahman et al. demonstrated the potential of using transfer learning with various pre-trained CNN models, such as AlexNet, ResNet18, DenseNet201, and SqueezeNet, to automatically detect bacterial and viral pneumonia using digital x-ray images. Their methodology, which involved training on a dataset of 5247 chest X-ray images, achieved classification accuracies of 98\%, 95\%, and 93.3\% for normal vs. pneumonia, bacterial vs. viral pneumonia, and a three-class classification scheme, respectively, showcasing the potential of deep learning in enhancing diagnostic accuracy \cite{rahman2020transfer}.

In the specific context of the COVID-19 pandemic, the challenges associated with traditional testing methods, such as RT-PCR, including shortages of test kits and delays in result delivery, have necessitated the exploration of alternative diagnostic methodologies. Deep learning frameworks utilizing chest X-rays have been proposed as a viable solution to facilitate the rapid triaging of COVID-19 patients. Singh et al. proposed a deep learning-based solution that employs a modified stacked ensemble model consisting of four CNN base-learners and Naive Bayes as a meta-learner to classify chest X-rays into COVID-19, pneumonia, and normal classes, achieving an accuracy of 98.67\% \cite{singh2020covidscreen}.

Moreover, the application of CNNs extends to the analysis of chest CT scans for the diagnosis of COVID-19. Yousefzadeh et al. introduced "ai-corona," a deep learning framework designed to assist radiologists in diagnosing COVID-19 by identifying abnormalities in chest CT scans. The framework employs a CNN-based feature extractor, coupled with average pooling and a fully-connected layer, to classify CT scans into COVID-19 abnormal, non COVID-19 abnormal, and normal classes, demonstrating the potential of CNNs in providing supplementary diagnostic insights to radiologists \cite{yousefzadeh2020aicorona}.

Despite these advancements, there remain challenges and opportunities for enhancement in the accuracy, reliability, and generalizability of CNN models in diagnosing various chest infections across multiple classes. This research endeavors to explore the development, optimization, and evaluation of a Deep Residual CNN for the multi-class diagnosis of chest infections, with a focus on providing a comprehensive exploration from model architecture to implementation. The ultimate goal is to contribute to the ongoing efforts in leveraging deep learning for enhanced medical diagnosis, particularly in the context of respiratory infections, and to provide a robust, reliable tool to assist healthcare professionals in the timely and accurate diagnosis of such conditions.

\section{Literature Review}
The integration of deep learning, particularly utilizing Convolutional Neural Networks (CNNs), into the realm of medical imaging, has been a subject of extensive research, especially in the context of diagnosing respiratory infections through chest X-ray images. The criticality of this application has been further underscored by the global COVID-19 pandemic, which has necessitated the development of rapid, accurate, and scalable diagnostic solutions.

One of the notable approaches in leveraging deep learning for the classification of COVID-19 chest X-ray images is the Decompose, Transfer, and Compose (DeTraC) deep CNN, proposed by Abbas et al. \cite{abbas2020detrac}. DeTraC addresses the challenge posed by the limited availability of annotated medical images by employing a class decomposition mechanism, which investigates class boundaries. The model demonstrated a high accuracy of 93.1\% (with a sensitivity of 100\%) in detecting COVID-19 cases from a comprehensive image dataset collected from several hospitals worldwide.

In another study by Sethi et al., four different deep CNN architectures were investigated for the diagnosis of COVID-19 using chest X-ray images \cite{sethi2020deep}. The models, pre-trained on the ImageNet database, demonstrated the potential of CNN-based architectures in the diagnosis of COVID-19, thereby providing a viable alternative to traditional testing methods, such as RT-PCR, which have been plagued by challenges such as false test results and test kit shortages.

Benbrahim et al. explored the application of Deep Transfer Learning (DTL) using CNN-based models InceptionV3 and ResNet50, combined with the Apache Spark framework, for classifying COVID-19 in chest X-ray images \cite{benbrahim2020deep}. The models achieved high accuracy in detecting COVID-19 X-ray images, with 99.01\% by the pre-trained InceptionV3 model and 98.03\% for the ResNet50 model, showcasing the efficacy of employing DTL in medical image classification.

Moreover, the application of CNNs is not limited to the diagnosis of COVID-19 but extends to other forms of pneumonia. Rajaraman et al. highlighted the advantages of visualizing and explaining the activations and predictions of CNNs applied to the challenge of pneumonia detection in pediatric chest radiographs \cite{rajaraman2019visualizing}. The study emphasized the importance of transparency and explainability in the learned behavior of CNNs, particularly in medical screening and diagnosis, to ensure that model predictions are interpretable and justifiable.

These studies underscore the potential and challenges of employing deep learning, particularly CNNs, in the diagnosis of respiratory infections through chest X-ray images. The exploration of various architectures, the application of transfer learning, and the emphasis on model explainability and interpretability are critical aspects that inform the ongoing research and development in this domain.

\begin{table*}[htbp!]
  \centering
  \begin{tabular}{cccc}  
    \hline 
    Class & Chest X-Ray Image & Total \# of the dataset \cite{baral2021covid} & Total \# of the dataset \cite{patel2020chest} \\
    \hline
    COVID-19 & \raisebox{-0.5\height}{\includegraphics[width=0.2\linewidth]{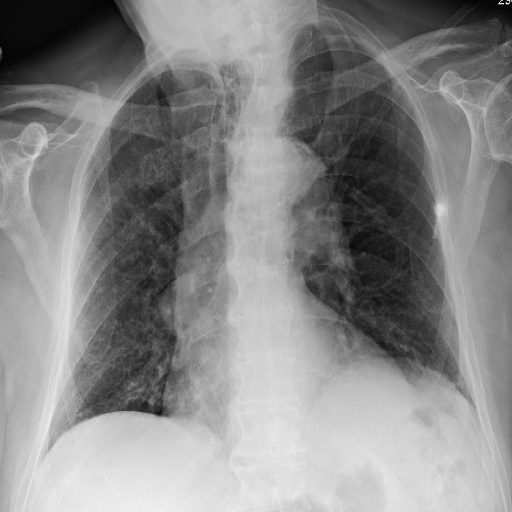}} & 3616 & 460 \tabularnewline
    \hline
    Fibrosis & \raisebox{-0.5\height}{\includegraphics[width=0.2\linewidth]{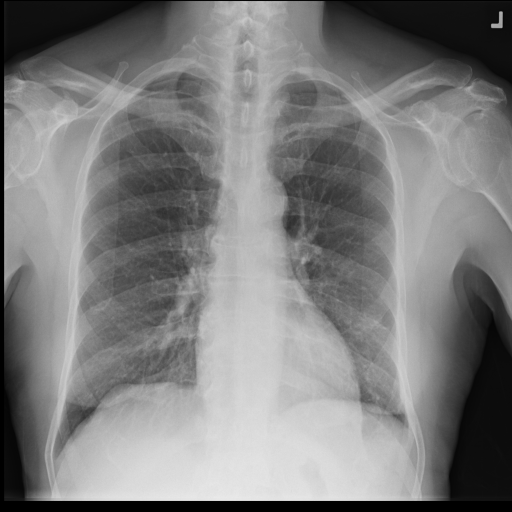}} & 1686 & - \\
    \hline
    Normal & \raisebox{-0.5\height}{\includegraphics[width=0.2\linewidth]{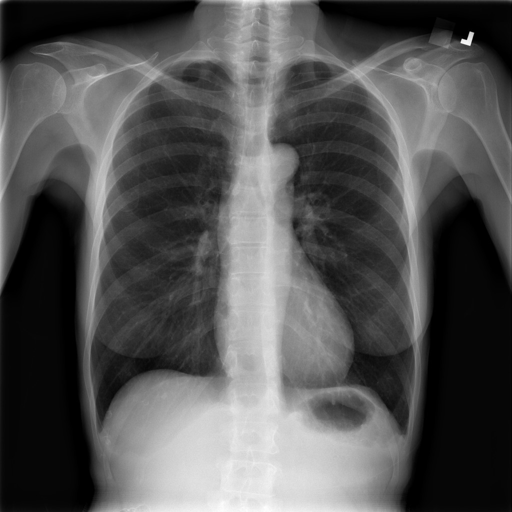}} & 11767 & 1266 \\
    \hline
    Pneumonia & \raisebox{-0.5\height}{\includegraphics[width=0.2\linewidth]{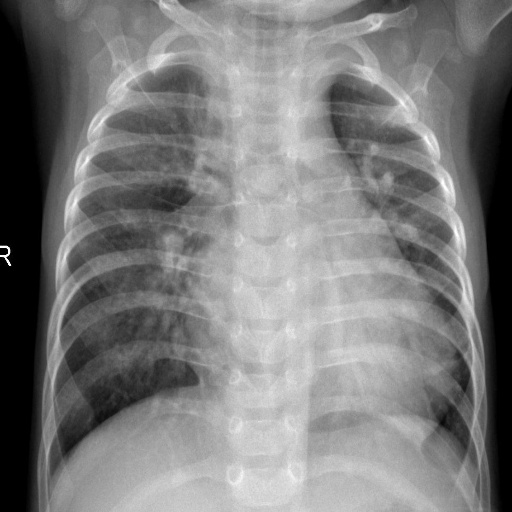}} & 4265 & 3418 \\
    \hline
    Tuberculosis & \raisebox{-0.5\height}{\includegraphics[width=0.2\linewidth]{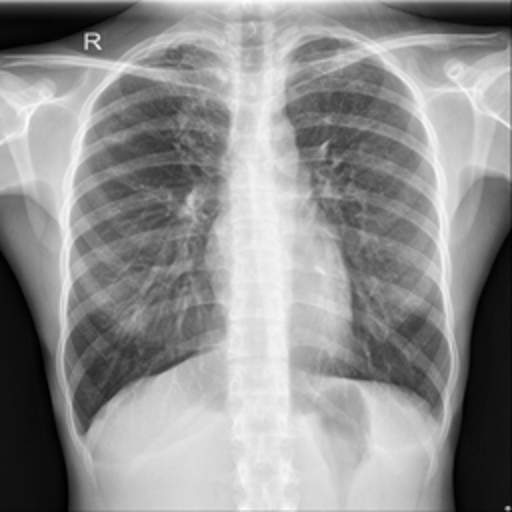}} & 3500 & - \\
    \hline
  \end{tabular}
  \caption{Image count of each classes}
  \label{tbl:datasets} 
\end{table*}

\section{Methodology}
The methodology employed in this research revolves around the application of a deep convolutional neural network (CNN) for the classification of chest X-ray images into five categories: COVID-19, Fibrosis, Normal, Pneumonia, and Tuberculosis. The dataset utilized for training and validating the model is derived from two sources: "Covid-19 Detection" \cite{baral2021covid} and "Chest X-ray (Covid-19 \& pneumonia)" \cite{patel2020chest}, which collectively provide a diverse and substantial pool of images across the aforementioned categories. The precise distribution of images across these categories is detailed in Table \ref{tbl:datasets}.  

\begin{figure}[htbp!]
    \centering
    \includegraphics[width=0.5\linewidth]{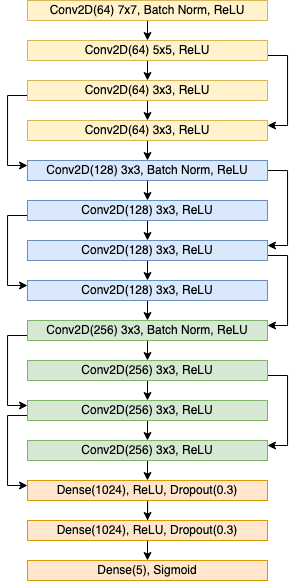}
    \caption{Model Architecture}
    \label{fig:model-arch}
\end{figure}

The CNN model developed for this research is structured to navigate through the intricacies and nuances of medical image classification, ensuring that it can effectively discern and differentiate between the various classes in the context of chest infections. The architecture of a CNN, shown in Figure \ref{fig:model-arch}, consists of several convolutional layers, each followed by a rectified linear unit (ReLU) activation function, and several layers fully connected towards the output. The convolutional layers are designed to extract hierarchical features from the input images, while the fully connected layers facilitate the final classification.

The model architecture begins with a convolutional layer with 64 filters, followed by batch normalization and ReLU activation. Subsequent convolutional layers are structured with varying numbers of filters and kernel sizes, strategically designed to progressively extract more complex features from the input images. Residual connections are introduced to preserve the gradient flow through the network, which is particularly crucial given the depth of the architecture. The final layers of the network are fully connected, culminating in a five-unit output layer with a sigmoid activation function, corresponding to the five classes of chest conditions.

The training process involves the use of the CrossEntropyLoss loss function and the Stochastic Gradient Descent (SGD) optimizer, with a learning rate of 0.001 and momentum of 0.9. The model is trained for a maximum of 50 epochs, with early stopping implemented to halt training if the validation loss does not improve after three consecutive epochs, thereby preventing overfitting and ensuring computational efficiency.

Model evaluation is conducted using a separate test dataset, ensuring that the assessment of the model’s performance is unbiased and representative of its ability to generalize to unseen data. Furthermore, additional evaluation metrics, such as a confusion matrix and classification report, are utilized to provide a comprehensive overview of the model's performance across all classes.

\section{Implementation}
The implementation of the deep residual CNN for multi-class chest infection diagnosis is executed using the PyTorch deep learning framework, renowned for its flexibility and efficiency in developing deep learning models \cite{paszke2019pytorch}. The dataset, sourced from \cite{baral2021covid} and \cite{patel2020chest}, is preprocessed and loaded using the torchvision package, ensuring that the images are resized to 224x224 pixels, converted to grayscale, and transformed into tensors, facilitating compatibility with the CNN.

The CNN architecture, as detailed in the methodology, is implemented with a series of convolutional layers, each followed by ReLU activation functions. The convolutional layers are designed to progressively extract hierarchical features from the input images, with the initial layers focusing on low-level features, such as edges and textures, and subsequent layers capturing more complex, high-level features. Residual connections are introduced to mitigate the vanishing gradient problem, ensuring that the gradient can propagate effectively through the network during the training process. The final layers of the network are fully connected, facilitating the classification of the extracted features into one of the five target classes.

The training process is executed over a maximum of 50 epochs, with the model being trained on a training dataset and validated on a separate validation dataset. The CrossEntropyLoss loss function and the SGD optimizer are utilized to update the model parameters during training, with a learning rate of 0.001 and momentum of 0.9. Early stopping is implemented to monitor the validation loss during training, halting the process if no improvement is observed after three consecutive epochs, thereby preventing overfitting and optimizing computational efficiency.

Model evaluation is conducted on a separate test dataset, ensuring an unbiased assessment of the model’s performance. The model’s predictions are compared with the actual labels to compute the accuracy of the model on the test data. Furthermore, additional evaluation metrics, such as the confusion matrix and classification report, are computed to provide a comprehensive overview of the model’s performance across all classes.

\section{Experimentation and Results}

\begin{table}[h!]
    \centering
    \begin{tabular}{ccccc}
        \hline
         & precision & recall & f1-score & support \\
        \hline
        0 & 0.94 & 0.96 & 0.95 & 1276 \\
        1 & 0.71 & 0.22 & 0.34 & 521 \\
        2 & 0.90 & 0.97 & 0.93 & 3974 \\
        3 & 0.99 & 0.99 & 0.99 & 2565 \\
        4 & 0.97 & 0.98 & 0.97 & 1043 \\
        \hline
        accuracy & - & - & 0.93 & 9379 \\
        macro avg & 0.90 & 0.82 & 0.84 & 9379 \\
        weighted avg & 0.93 & 0.93 & 0.92 & 9379 \\
        \hline
    \end{tabular}
    \caption{Classification Report: Precision, Recall, and F1 Score}
    \label{tab:result}
\end{table}

The experimentation phase was meticulously conducted to validate the efficacy of the implemented deep residual CNN in classifying chest X-ray images into five distinct categories: COVID-19, Fibrosis, Normal, Pneumonia, and Tuberculosis. The model was subjected to a series of tests using a dedicated test dataset, which was not exposed to the model during the training phase, ensuring an unbiased evaluation of its performance.

The results of the experimentation are encapsulated in the classification report, which provides a detailed overview of the model’s performance across various metrics, including precision, recall, and F1-score, for each class, as well as the overall accuracy of the model. The model achieved an overall accuracy of 93\%, demonstrating a high degree of proficiency in classifying the chest X-ray images across the five target classes.

In terms of individual class performance, the model exhibited varying degrees of success. For Class 0 (COVID-19), the model achieved a precision of 94\%, recall of 96\%, and F1-score of 95\%. Class 1 (Fibrosis) presented a notable challenge, with the model achieving a precision of 71\%, but a significantly lower recall of 22\% and F1-score of 34\%. Class 2 (Normal) was classified with a precision of 90\%, recall of 97\%, and F1-score of 93\%. Class 3 (Pneumonia) and Class 4 (Tuberculosis) were classified with high proficiency, achieving precision, recall, and F1-scores of 99\%, 99\%, 99\% and 97\%, 98\%, 97\% respectively.

\begin{figure}[htbp!]
    \centering
    \includegraphics[width=1.0\linewidth]{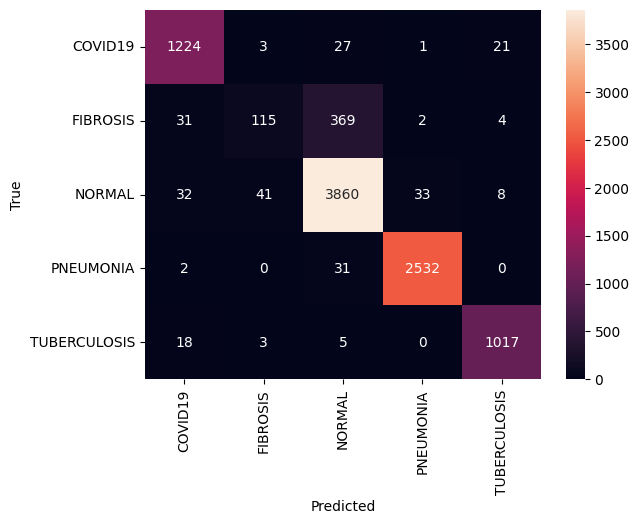}
    \caption{Confusion Matrix and its Interpretation}
    \label{fig:confusion-matrix}
\end{figure}

The model demonstrated a particularly high proficiency in classifying Pneumonia and Tuberculosis, which can be attributed to the distinctive visual features associated with these conditions in chest X-ray images, which are effectively captured and classified by the CNN. However, the classification of Fibrosis presented a notable challenge, potentially due to the subtle and nuanced visual features associated with this condition, which may be less distinguishable to the model, thereby necessitating further investigation and optimization.

The confusion matrix, which provides a detailed overvi\-ew of the model’s classifications, further underscores the challenges and successes encountered by the model across the various classes. The matrix reveals that while the model is highly proficient in classifying certain conditions, misclassifications do occur, particularly between classes that may share similar visual features in the X-ray images.

\section{Discussion}
The exploration into the realm of multi-class chest infection diagnosis via deep residual CNNs has unveiled a spectrum of insights and potential avenues for future research. The implemented model, which was trained and validated using a dataset amalgamated from \cite{baral2021covid} and \cite{patel2020chest}, demonstrated a commendable overall accuracy of 93\%. However, the nuanced disparities in the model’s performance across the different classes warrant a deeper exploration and discussion.

The model exhibited a particularly proficient capability in classifying certain conditions, such as Pneumonia and Tuberculosis, which can be attributed to the distinctive and pronounced visual features associated with these conditions in chest X-ray images. This aligns with existing literature that highlights the success of deep learning models in identifying patterns and features in medical images that are often imperceptible to the human eye \cite{litjens2017survey}. However, the classification of Fibrosis posed a notable challenge, as evidenced by the lower recall and F1-score. This could potentially be attributed to the subtle and often nuanced visual features associated with Fibrosis in chest X-ray images, which may be less distinguishable and discernible to the model.

The implementation of residual connections within the CNN architecture was instrumental in mitigating the vanishing gradient problem, facilitating the effective propagation of gradients through the network during the training process \cite{he2016deep}. This architectural choice was pivotal in enabling the model to learn and extract hierarchical features from the input images, contributing to its overall proficiency in classifying the images across the various classes. However, the disparities in performance across the classes suggest that further optimization and refinement of the model architecture and training process are warranted.

The challenge posed by the classification of Fibrosis underscores the importance of further research and development in enhancing the model’s capability to discern and differentiate between subtle and nuanced visual features in chest X-ray images. This could potentially involve the exploration and implementation of more sophisticated and advanced CNN architectures, as well as the incorporation of additional data augmentation techniques during the training process to enhance the model’s generalization capability \cite{shorten2019survey}. Furthermore, the exploration into transfer learning, leveraging pre-trained models on larger and more diverse datasets, could potentially enhance the model’s feature extraction capabilities and overall classification proficiency \cite{tan2018survey}.

In conclusion, while the implemented deep residual CNN demonstrated a commendable capability in classifying chest X-ray images across various conditions, the insights derived from the experimentation and results pave the way for future research and development. Focusing on enhancing the model’s proficiency in classifying conditions that present more subtle and nuanced visual features in the images will be pivotal in advancing the field of automated medical image diagnosis and providing valuable tools and resources in the realm of medical diagnostics.

\section{Conclusion and Future Work}
The journey through developing and evaluating a deep residual CNN for multi-class chest infection diagnosis has been both insightful and revealing. The model, trained and validated on a dataset amalgamated from \cite{baral2021covid} and \cite{patel2020chest}, demonstrated a robust capability in classifying chest X-ray images across various conditions, achieving an overall accuracy of 93\%. However, the nuanced disparities in performance across different classes, particularly the challenges encountered in the classification of Fibrosis, underscore the complexity and multifaceted nature of automated medical image diagnosis.

The implemented model leveraged the power of deep learning, particularly the efficacy of CNNs in image classification tasks, to extract and learn hierarchical features from the input images, contributing to its proficiency in classifying the images across the target classes \cite{lecun2015deep}. The incorporation of residual connections within the architecture facilitated the effective training of the network, mitigating the vanishing gradient problem and enabling the model to learn more complex and nuanced features from the images \cite{he2016deep}. However, the challenges encountered in the classification of certain conditions highlight the limitations and areas requiring further research and optimization.

Looking forward, the insights and findings derived from this research pave the way for several avenues of future work. Firstly, the exploration and implementation of more advanced and sophisticated CNN architectures, potentially leveraging the capabilities of architectures that have demonstrated proficiency in similar tasks, could enhance the model’s feature extraction and classification capabilities \cite{tan2018survey}. Additionally, the incorporation of additional data, particularly for classes that presented challenges in classification, could enhance the model’s training and generalization capabilities, providing a more robust and comprehensive understanding of the various conditions.

Furthermore, the exploration into additional data augmentation techniques, as well as the implementation of transfer learning leveraging pre-trained models on larger and more diverse datasets, could potentially enhance the model’s feature extraction capabilities and overall classification proficiency \cite{shorten2019survey}. The implementation of more sophisticated training and optimization techniques, potentially exploring alternative loss functions and optimization algorithms, could further enhance the model’s training process and overall performance.

In conclusion, while the implemented model demonstrated a commendable capability in classifying chest X-ray images, the insights derived from the experimentation and results provide a foundation upon which future research and development can build. Enhancing the model’s proficiency in classifying conditions that present more subtle and nuanced visual features in the images, as well as optimizing and refining the model architecture and training process, will be pivotal in advancing the field and providing valuable tools and resources in the realm of medical diagnostics.

\bibliography{references}
\bibliographystyle{elsarticle-num}

\end{document}